\documentstyle[aps]{revtex}

\begin{document} \draft

\preprint{HEP/123-qed}

\title{Field theory of avalanche formation}
\author{Alexander I. Olemskoi, Alexei V. Khomenko \\
{\it Sumy State
University}\\ {\it 2, Rimskii-Korsakov St., 40007 Sumy, Ukraine}\\
{\it E-mail: olemskoi@ssu.sumy.ua}  }
\date{\today}

\maketitle

\begin{abstract}
Self-organizing system is studied whose behavior is governed by  field of an
order parameter, a fluctuation amplitude of conjugate field
and a couple of Grassmannian conjugated fields that define the entropy as
a control parameter. Within the framework of self-consistent approach
the dependencies of macro- and microscopic susceptibilities as well
as memory and nonergodicity parameters are determined as a functions
of the intensities of thermal and quenched disorders. Making use of the
sandpile model shows that proposed scheme determines the
conditions of avalanches formation in self-organized criticality
phenomena.

\end{abstract}


\newpage

\section{Introduction}\label{sec:level1}

Nowadays there is such an original situation in the theory of
self-organizing systems. On the one hand, the synergetic concept has
been developing successfully for more than twenty years. It allows us
to explain the self-organization (ordering) of open system subjected to
the environment disorder \cite{1}.  On the other hand, the
phenomena of the self-organized criticality such as an avalanche
motion of the sand grains on inclined surface (sandpile model)
\cite{2}, intermittency in biological evolution \cite{3}, earthquakes
and forest fire, pinning in the random medium etc. (see
\cite{4}) have been actively investigated for about ten years.
However, in spite of the fact that the mission of both synergetics
and theory of self-organized criticality is to explain the same
phenomenon -- the self-organization, they develop independently.  It
became so as within the framework of the synergetic approach the
single statistical ensemble (formation of an avalanche) is
investigated, whereas the examination of self-organized criticality
models \cite{2}-\cite{4} is reduced usually to study of evolution of
hierarchical avalanche ensemble \cite{5} and is based
on numerical methods and scaling representations \cite{6}.

Being the object of synergetics the formation of a single avalanche
still remains an open question. It is evident that
its reviewing is the only way to set connection between specified
directions. Since phenomena of self-organized criticality are
caused by avalanche ensemble behavior, and certain closed
region of the state space corresponds to each avalanche, the
standard formulation of the synergetic problem requires a nontrivial
extension -- while ordering self-organizing system we have to
describe not only a symmetry breaking, but the  ergodicity breaking
that induces the clusterisation of phase space as well. In our paper
a solution of this problem is suggested. In so doing we will
investigate only the self-organization picture digressing from
avalanche ensemble consideration (see \cite {7}).

The first attempt of field description of an avalanche formation was
undertaken within the framework of the one-parameter approach \cite
{8}, based on nonlinear diffusion equation -- thus, the feedback
between open system and environment was not taken into
account. Recently, two-parameter models were suggested, where the
environment is represented either by control parameter \cite{9}, or
conjugate field \cite{10}.  According to \cite{9} avalanche-like
subcritical mode is formed in case of adiabatic relationship of
characteristic times and first-order transition mechanism. The
approach of Ref.~\cite{10} allowed us to determine the critical indexes
representing a scaling behavior of a self-organizing system within
the framework of  mean field approximation. Below we propose the
generalized self-consistent
scheme that is taken into account total number of the freedom degrees.
Due to this we obtain not only
mutually supplementary results \cite{9,10} but also the complete
analytical description of an avalanche formation.

The suggested approach is based on synergetic generalization of the
thermodynamic theory of phase transitions.  The main feature of this
theory is that within the conserved system (thermostat) a subsystem is
segregated to represent a hydrodynamical mode whose amplitude
is qualified as the order parameter \cite{11} that determines the state of the subsystem. Here it is assumed
that thermostat influences upon value of the order parameter $\eta$
both thermally -- by varying of a control parameter $S$, and
immediately -- by varying the field $h$ conjugated to the order
parameter (in case of magnet the values $ \eta $, $h$, $S$ mean the
magnetization, the magnetic field and the entropy). The distinctive
peculiarity of the thermodynamic approach is that a postulation of
the one-sided influence of thermostat on the ordering subsystem is
declared, but not the reverse -- the order parameter $\eta$
variation does not influence upon thermostat state parameters $h$ and
$S$. The synergetic approach considers the connection between an
open subsystem and thermostat to be two-sided, so that the control
parameter $S$ and conjugate field $h$ turn out to be the functions of
order parameter $\eta$. This connection manifests itself especially
while the phase transition is described kineticaly. So, the standard
picture meeting the  Landau-Khalatnikov dissipative dynamics is
realized in adiabatic approximation, when the relaxation time of the
order parameter is much longer that corresponding times for conjugate
field and control parameter \cite {12}.  To this end the making use
of Lorentz system, primarily suggested in order to describe the
turbulent airflows \cite{13}, happens to be rather convenient.

Our paper is organized as follows.
In sections\ \ref{sec:level2} (basic expressions), \ \ref{sec:level3} 
(Lagrange formalism), and \ \ref{sec:level7} 
(Appendix) is shown that the Lorentz system corresponds to
the simplest Lagrangian of the supersymmetric field with components
giving an order parameter $\eta$, a conjugate field $h$, and entropy $S$.
It is rather important
that in this case a combination of the Grassmannian components of
superfield plays the role of control parameter $S$ in
contradistinction to the usual field theory of a stochastic system
\cite{14}, where they are an auxiliary variables which have no
physical meaning.  The fact that the variables $\eta$, $h$, $S$ meet
the vector of supersymmetric space is a reflection of the
self-consistent behavior of the synergetic system (in contrast to the
statistical field scheme \cite {14}, where the superfield is only a
convenient technical method). The study of superfield correlators
is carried out in section \ \ref{sec:level4}.
As is known the components of such correlators are not independent in
ergodic state -- the presence of supersymmetry causes the
fluctuation-dissipation theorem which connects
stated components \cite{14}. If the quenched disorder is appeared, an
ergodicity loosing happens that breaks the supersymmetry in
its turn.  This leads to appearance of singular additives to
correlators that define memory and nonergodicity parameters $q$,
$\Delta$. The basic result of  our work is the defining quantities
$q$ and $\Delta$ dependencies on thermal and quenched disorders
intensities.  This allows us to find the conditions of avalanches
formation in the self-organized criticality mode appearance.
The generic example of such a process is considered in 
section\ \ref{sec:level5} as 
the flow of sand on inclined surface. It is shown that this process
can be represented by the Lorentz system, considering
the horizontal and vertical components of the sand grain velocities to be
the order parameter $\eta$ and conjugate field $h$, and the tangent
of a surface inclination angle as the control parameter $S$.
Final section\ \ref{sec:level6} is devoted to discussion of obtained results. 
It is shown
that a parameter which determines the transition to time irreversible
regime is given by ratio of a time of the quantum fluctuation to the one of
order parameter. A critical value of effective interaction that bounds the
domain of ordered state is found.

\section{Lorentz system}\label{sec:level2} 

To expound the
microscopic scheme of the self-organization description
let us study firstly the system consisting of Bozon and Fermion
gases whose interaction is characterized by  potential $v$.
Within the framework of secondary quantization Bozons are described
by the $b_l^{+}$ and $b_l$ operators, satisfying the usual
commutation relation:  $[b_l,b_m^{+}]=\delta_{lm}$, where $l,m$ are
the site numbers.  The two-level Fermion subsystem is represented by
operators $a_{l\alpha}^{+}, a_{l\alpha}$, $\alpha=1,~ 2$, for which
the anti-commutation relation $\{a_{l\alpha},
a_{m\beta}^{+}\}=\delta_{lm}\delta_{\alpha\beta}$ is fulfilled.  The
occupation numbers $b_{\rm{\bf k}}^{+}b_{\rm{\bf k}}$ determine the
Bozon distribution within $\rm{\bf k}$-representation that corresponds
to the Fourier transform over lattice sites $l$. To represent the
Fermi subsystem we should introduce the operator $d_{l}\equiv
a_{l1}^{+}a_{l2}$ determining the polarization with respect
to the saturation over levels $\alpha=1,~2$, as an
addition to the occupation numbers $n_{l\alpha}\equiv
{a_{l\alpha}^{+}a_{l\alpha}}$.  As a result, behavior of the
system under consideration is defined by Dicke Hamiltonian
\begin{equation} H=\sum_{\rm{\bf k}} \left\{(E_{1}n_{{\rm{\bf k}}1}+
E_{2}n_{\rm{\bf k}2})+\omega_{\rm{\bf k}}b_{\rm{\bf k}}^{+}b_{\rm{\bf
k}} +{{\rm i} \over 2} v({b_{\rm{\bf k}}^{+}}d_{\rm{\bf
k}}-d_{\rm{\bf k}}^{+}b_{\rm{\bf k}}) \right\}, \label{1}
\end{equation} where the $\rm{\bf k}$-representation is used,
$E_{1,2}$ are the Fermi levels energies, $\omega_{\rm{\bf k}}$
represents the Bozon dispersion law and the imaginary unit before
the interaction term $v$ reflects the Hermitity property,
the Planck constant is $\hbar=1$.

The Heisenberg equations of motion corresponding to Hamiltonian
(\ref{1}) have the form 

\def\theequation{\arabic{equation}{a}} \setcounter{equation}{1} 
\begin{equation} 
\dot{b}_{\rm{\bf k}}=-{\rm
i}\omega_{\rm{\bf k}}b_{\rm{\bf k}}+(v/2)d_{\rm{\bf k}}, \label{2a}
\end{equation}
\def\theequation{\arabic{equation}{b}} \setcounter{equation}{1}
\begin{equation}
\dot{d}_{\rm{\bf k}}=-{\rm i}\Delta d_{\rm{\bf k}}+(v/2)b_{\rm{\bf
k}}(n_{\rm{\bf k}2}-n_{\rm{\bf k}1}), \label{2b}
\end{equation}
\def\theequation{\arabic{equation}{c}} \setcounter{equation}{1}
\begin{equation}
\dot{n}_{\rm{\bf
k}1}=(v/2)(b_{\rm{\bf k}}^{+} d_{\rm{\bf k}}+d_{\rm{\bf k}}^{+}
b_{\rm{\bf k}}), \label{2c}
\end{equation}
\def\theequation{\arabic{equation}{d}} \setcounter{equation}{1}
\begin{equation}
\dot{n}_{\rm{\bf
k}2}=-(v/2)(b_{\rm{\bf k}}^{+} d_{\rm{\bf k}}+d_{\rm{\bf k}}^{+}
b_{\rm{\bf k}}), \label{2d} \end{equation} 
\def\theequation{\arabic{equation}}\setcounter{equation}{2}

\noindent 
where the dot stands for a
derivative with respect to time and the quantity $\Delta\equiv
{E_{2}-E_{1}}$ is introduced. In case of
resonance suggested the first terms in the right-hand
sides of  equations (\ref{2a}), (\ref{2b}) that contains frequencies
$\omega_{\rm{\bf k}}$ and $\Delta$ may be eliminated  by
extracting the multipliers $\exp(-{\rm i}\omega_{\rm{\bf k}} t)$ and
$\exp(-{\rm i}\Delta t)$ in the time dependencies $b_{\rm{\bf
k}}(t)$, $d_{\rm{\bf k}}(t)$, respectively. On the other hand, if the
dissipation is taken into account, these frequencies obtain imaginary
additions $-{\rm i}/\tau_{\eta}$, $-{\rm i}/\tau_{h}$ characterized
by relaxation times $\tau_{\eta}$, $\tau_{h}$ (here the conditions
${\rm Im}~\omega_{\rm{\bf k}}<0$, ${\rm Im}~\Delta<0$ reflect the
"causality" principle). As a result, equations (\ref{2a}), (\ref{2b})
get the dissipative terms $-b_{\rm{\bf k}}/{\tau_{\eta}}$,
$-d_{\rm{\bf k}}/{\tau_h}$, where $\tau_{\eta}$ is the relaxation
time of Bozon distribution and $\tau_{h}$ is the Fermion
polarization time.  One can suppose that the dissipation influences
onto the Fermi levels occupancies $n_{\rm{\bf k}\alpha}(t)$ also.
However, since the stationary values $n_{\rm{\bf k}\alpha}^0\not=0$
(and in case of external pumping $n_{\rm{\bf k}2}^0>n_{\rm{\bf
k}1}^0$) the dissipative terms in Eqs.~(\ref{2c}), (\ref{2d}) have
much complicated form: $\ -(n_{\rm{\bf k}\alpha}-n^0_{\rm{\bf
k}\alpha})/\tau_S$, $\alpha=1,~2$, where $\tau_S$ is the relaxation
time of the Fermion distribution over level.

Now, let us introduce the macroscopic quantities:
\begin{eqnarray}
&& \eta_{\bf k} \equiv \langle b_{\bf k}^{+}\rangle=
\langle b_{\bf k}\rangle, \qquad
h_{\bf k} \equiv \langle d_{\bf
k}\rangle=\langle d_{\bf k}^{+}\rangle, \nonumber\\
&& S_{\bf k} \equiv \langle n_{{\bf k}2}-n_{{\bf k}1}\rangle,\qquad
S^0_{\bf k} \equiv \langle n_{{\bf k}2}^0-n_{{\bf k}1}^0
\rangle, \label{3} \end{eqnarray}
where the angular brackets mean thermodynamic averaging.
Neglecting the correlation in distribution of particles over quantum
states the Heisenberg equations (2), being contemplated by
dissipative terms, result in the Lorentz system  

\def\theequation{\arabic{equation}{a}} \setcounter{equation}{3} 
\begin{equation}
\tau_{\eta}\dot{\eta}=-\eta+A_{\eta}h, \label{4a}
\end{equation}
\def\theequation{\arabic{equation}{b}} \setcounter{equation}{3}
\begin{equation}
\tau_{h}\dot{h}=-h+A_{h}\eta S, \label{4b}
\end{equation}
\def\theequation{\arabic{equation}{c}} \setcounter{equation}{3}
\begin{equation}
\tau_{S}\dot{S}=(S^{0}-S)-A_{S}\eta h.  \label{4c} 
\end{equation} 
\def\theequation{\arabic{equation}}\setcounter{equation}{4} 

\noindent
Here in terms of one-mode approximation
the dependence on the wave vector ${\rm{\bf k}}$ is omitted and the
constants defined by relationships $2A_{\eta}\equiv v\tau_{\eta}$,
$2A_{h}\equiv v\tau_{h}$, $A_{S}\equiv 2v\tau_{S}$ are introduced.
Equations (4) contain following seven constants -- pumping
parameter $S^{0}$, three relaxation times $\tau$ and three coupling
constants $A$. But as there are above relations between these last
ones caused by interaction parameter $v>0$, only five of them are
independent.  Since four of these fix the scales
for quantities $\eta$, $h$, $S$, $t$, so only the parameter of
thermal disorder $S^{0}$ plays a substantial role and its value
determines only the system behavior \cite{1,12}.

To analyze equations (4) we introduce the scales
$\eta_{m},~h_{m},~S_{c}$ defining ranges of  the variation for the
order parameter $\eta$, the conjugate field $h$ and the control
parameter $S$:  \begin{eqnarray} && \eta_{m}^{-2}\equiv
A_{h}A_{S}=\tau_{h}\tau_{S}v^2, \qquad h_m^{-1}\equiv
A_{\eta}/\eta_m=(\tau_{\eta}/2)(\tau_{h}\tau_S)^{1/2}v^2;\nonumber\\
&& S_c^{-1}\equiv A_{\eta}A_h=2^{-2}\tau_{\eta}\tau_{h}v^2.\label{5}
\end{eqnarray}
Then, using  magnitudes $\eta, h, S$ normalized by $\eta_m,~h_m,~S_c$
values, we results Eqs.~(4) in the form:  

\def\theequation{\arabic{equation}{a}} \setcounter{equation}{5} 
\begin{equation}
\tau_{\eta}\dot{\eta}=-\eta+h, \label{6a}
\end{equation}
\def\theequation{\arabic{equation}{b}} \setcounter{equation}{5}
\begin{equation}
\tau_{h}\dot{h}=-h+\eta S,\label{6b}
\end{equation}
\def\theequation{\arabic{equation}{c}} \setcounter{equation}{5}
\begin{equation}
\tau_{S}\dot{S}=(S^{0}-S)-\eta h.\label{6c}
\end{equation} \def\theequation{\arabic{equation}}\setcounter{equation}{6}

\noindent 
In terms of adiabatic approximation
$\tau_h,\tau_S\ll\tau_{\eta}$ the left-hand sides of equations
(\ref{6b}), (\ref{6c}) may be set equal to zero. Thus, we derive to a
result \begin{equation} h=S^{0}\eta/({1+\eta^{2}}),\qquad
S=S^{0}/({1+\eta^{2}}).  \label{7} \end{equation} With the order
parameter growth in physical domain $\eta \in [0, 1]$ the conjugate
field increases and the control parameter decreases monotonically. If
$\eta>1$ the $h(\eta)$ dependence is of decreasing shape, that
corresponds to unstable state.

Inserting (\ref{7}) into (\ref{6a}) we obtain the Landau-Khalatnikov
equation \begin{equation} \tau_{\eta}\dot{\eta}=-\partial{V}/
\partial{\eta},\qquad V\equiv {1 \over
2}\left(\eta^{2}-S^{0}\ln(1+\eta^{2})\right).\label{8} \end{equation}
Its form is defined by synergetic potential $V(\eta)$ having its
minimum at point $\eta_{0}{=}(S^{0}{-}1)^{1/2}$.  Hence it is seen
that stationary value of the order parameter $\eta_{0}\not= 0$ is
realized under the thermal disorder conditions $S^{0}>1$
($S^{0}>S_{c}$ in usual units).  Thus, the magnitude $S_c$ defined in
last equality (\ref{5}) is the critical value of the control
parameter. According to Eq.~(\ref{7}) in the stationary state one has
$h_0=(S^{0}-1)^{1/2}$, $S_{0}=1$.  The last equality implies that
despite supercritical value $S^{0}>1$ of thermal disorder the system
relaxes so, that the stationary value of the control parameter
$S_{0}=1$ is reduced to the critical one.

The mentioned relaxation is provided by the negative feedback of
order parameter $\eta$ and conjugate field $h$ with control
parameter $S$ that is described by the last term of equation
(\ref{6c}).  This feedback, displaying the Le Chatelier principle for
the self-organizing system, compensates the $S(\eta)$, $h(\eta)$
thermostat's state parameters increase which takes place, when this
feedback is absent. On the other hand, the positive feedback of
quantities $\eta$ and $S$ with  $h$ in (\ref{6b}) is the reason for
the self-organization. It is obvious that the stationary state
$\eta_0$, $h_0$, $S_0$ can be realized only under the condition of
inverse subsystem influence, characterized by order parameter $\eta$, 
on the thermostat parameters $h, S$.  It is worthwhile to note that
inverting the signs of nonlinear terms in equations (6) causes the
minus appearance in the right-hand side of the first equality
(\ref{7}), so that the susceptibility $\chi={\rm d}\eta/{\rm d}h$
becomes negative and this case does not meet the stable state.

Thus, the self-organization process takes place only if the negative
feedback of order parameter $\eta$ and field $h$ with control parameter
$S$ and the positive feedback of $\eta$ and  $S$ with $h$ both exist.
According to (\ref{6b}),
(\ref{6c}) such a choice of signs is determined by the fact that the
negative feedback provides falling-down of the control parameter $S$ in the
course of time,
whereas the positive one insures the field $h$ growth.  Further we
shall show that the value $S$ is reduced to the entropy and its
decrease reflects the non-conservation of the self-organizing system
for which the second law of thermodynamics is not fulfilled. The
crucial role of the increasing character of  field $h$ is
stipulated by the fact that the linear equation ($6{\rm a}$)
for the order parameter $\eta$ contains
namely the field $h$. As a result,
influence of the field $h$ on the velocity of $\eta(t)$ increase, and
on the self-organization process also, is direct, whereas the
influence of control parameter is indirect.

The described scheme of self-organization meets the second-order
phase transition. To describe the first-order transition one ought to
set the relaxation time $\tau_\eta$ of order parameter to be the
function of its value $\eta$ \cite{1,12}. Such a scheme of an
avalanche formation is represented in \cite{7}.

\section{Lagrange formalism}\label{sec:level3}

In the previous section we omitted, within the adiabatic
approximation, the fluctuations of conjugate field $h$ and control
parameter $S$ and, thus, we made it possible to reduce the Lorentz
system (6) to the Landau-Khalatnikov equation (\ref{8}). To form
the Lagrange formalism one should accomplish the reverse transition
supposing a fluctuation source $\zeta$ appearance in (\ref{8}). If the
nonhomogeneity is considered, the generic expression is reduced to
the Langevine equation (see \cite{15} for example) \begin{equation}
\dot{\eta}({\rm{\bf r}},t)-{\nabla^{2}}\eta({\rm{\bf
r}},t)=f({\rm{\bf r}},t)+\zeta({\rm{\bf r}},t), \label{9}
\end{equation} where $\rm{\bf r}$ is the coordinate measured in units
of the correlation lengths $\xi$ and $t$ is the time related to the
scale $\tau_{\eta}$, the force $f=-{V_{0}}'(\eta)$, ${V_{0}}'\equiv
{\partial V_{0}/\partial \eta}$ is defined by dependence
$V_{0}(\eta)$ for the bare potential related to fluctuations
intensity $T$. The term $-{\nabla}^{2}\eta$ in the left-hand side of
Eq.~(\ref{9}) takes into account the spatial nonhomogeneity within
the framework of Ginzburg-Landau model. The expression (\ref{9}) is
valid for nonconserved order parameter, otherwise terms
$-{{\nabla}^{2}\eta}$, $f$ obtain the additional operator
$-{\nabla}^{2}$ \cite{16}.  The fluctuational term is normalized by
the white noise conditions \begin{equation} \langle\zeta({\rm{\bf
r}},t)\rangle =0,\qquad \langle\zeta({\rm{\bf r}},t)\zeta({\rm{\bf
r}}',t')\rangle=T\delta({\rm{\bf r}}-{\rm{\bf r}}')\delta(t-t'),
\label{10} \end{equation} which correspond to averaging over the
Gaussian distribution with dispersion $T$.

To construct the Lagrangian corresponding to the Langevine equation
(\ref{9}), let us use the standard field scheme \cite{14} based on a
generating functional \begin{equation} Z\{\eta({\rm{\bf
r}},t)\}=\left\langle\prod_{({\rm{\bf
r}},t)}\delta(\dot{\eta}-{\nabla}^{2}\eta-f- \zeta)\
\det\left|{\delta\zeta} \over {\delta\eta} \right |\right\rangle
\label{11} \end{equation} being the
generalization of the partition function.  Herein the continual
product of $\delta$-functions takes into account that the condition
(\ref{9}) ought to be satisfied with all values $\rm{\bf r}$, $t$ and
the determinant represents the Jacobian of $\zeta$ to $\eta$
transition. We apply the Fourier transform for $\delta$-function
leading to appearance of the field $\varphi({\rm{\bf r}},t)$ and
 introduce the Grassmannian conjugate fields $\psi({\rm{\bf r}},t)$,
$\bar\psi({\rm{\bf r}},t)$ into operation for integrated
representation of the determinant. Thus, the equality (\ref{11})
assumes the canonical form \begin{equation} Z\{\eta\}=\int
P\{\eta,\varphi,\bar\psi,\psi\}D\varphi D\bar\psi D\psi,\qquad
P\propto e^{-S}, \qquad S\equiv\int
{\cal L}(\eta,\varphi,\bar\psi,\psi){\rm d}{\rm{\bf r}}{\rm{d}}t,
\label{12} \end{equation} where the Lagrangian \begin{equation}
{\cal L}=\varphi(\dot{\eta}-{\nabla}^{2}\eta)+\bar\psi(\dot{\psi}-
{\nabla}^{2}\psi)-{\varphi}^{2}/2+ \varphi{V_{0}}'(\eta)+
{V_{0}}''(\eta){\bar\psi}\psi \label{13}
\end{equation}
is measured in noise intensity $T$ units.
The form of the corresponding Euler equations is as follows:

\def\theequation{\arabic{equation}{a}} \setcounter{equation}{13} 
\begin{equation}
\dot\eta-\nabla^2\eta=-{V_0}'(\eta)+\varphi,\label{14a}
\end{equation}
\def\theequation{\arabic{equation}{b}} \setcounter{equation}{13}
\begin{equation}
\dot\varphi+\nabla^2\varphi={V_0}''(\eta)\varphi+{V_0}'''(\eta) \bar\psi\psi, 
\label{14b}
\end{equation}
\def\theequation{\arabic{equation}{c}} \setcounter{equation}{13}
\begin{equation}
\dot\psi-\nabla^2\psi=-{V_0}''(\eta)\psi,\label{14c}
\end{equation}
\def\theequation{\arabic{equation}{d}} \setcounter{equation}{13}
\begin{equation}
\dot{\bar{\psi}}+\nabla^2\bar\psi={V_0}''(\eta)\bar\psi.\label{14d}
\end{equation} \def\theequation{\arabic{equation}}\setcounter{equation}{14}  

\noindent
The first one can be reduced to the
Langevine equation (\ref{9}) by replacing the field $\varphi$ by
stochastic component $\zeta$.  Since this equation corresponds to
the maximum of a probability $P$ distribution in (\ref{12}),
$\varphi$ represents the amplitude of the most probable fluctuation
of the field conjugated to the order parameter $\eta$ (the force $f$
is the average value of this field).  Obviously, the conditions
$\langle\zeta\rangle=0$, $\varphi\not= 0$ imply that during the
self-organization the bare
Gaussian distribution transforms from unimodal into bimodal
form with its maximums at  points $\pm\varphi$.  The signs distribution in front of the gradient
terms in Eqs.~(14) is quite remarkable:  albeit the standard
combination inherent in relaxation processes like
diffusion is realized for components $\eta$, $\psi$, but the fields
$\varphi$, $\bar\psi$ contain nonhomogeneity terms with opposite
signs that means the autocatalitical increase of those components. It
is shown in Appendix how the field equations (14) are reduced to the
Lorentz system (6).

The most elegant method to represent developed field scheme is
to incorporate the components $\eta$, $\psi$, $\bar\psi$ and the
generalized force $\phi\equiv -\delta V_0\{\eta \} /\delta \eta$ into
supersymmetrical field \begin{equation}
\Phi=\eta+\bar\psi\chi+\bar\chi\psi+\bar\chi\chi\phi, \label{15}
\end{equation} where Grassmannian coordinates $\chi$, $\bar\chi$
posses the same anti-commutation properties as the $\psi$,
$\bar\psi$ fields.  To represent the Lagrangian (\ref{13})
supersymmetricaly one should first of all replace the bare potential
$V_0(\eta)$ by renormalized one $\widetilde{V}(\eta)$, presented
in motion equation ~(\ref{A.7}),~\footnote{$^)$$\ $ It is shown in Appendix
that the renormalization is caused by self-consistency of the
superfield components (\ref{15}).}$^)$ and get rid of gradient
addends, committing transform to the variational derivatives
$\widetilde{V}'\{\eta\}\equiv\delta\widetilde{V}\{\eta\}/
{\delta\eta}={\partial\widetilde{V}(\eta)}/{\partial
\eta}-\nabla^2{\eta}$, $\widetilde{V}\{\eta\}\equiv\int
\widetilde{V}(\eta)\rm d{\rm{\bf r}}$.  Then, expressing fluctuation
amplitude $\varphi$ in terms of generalized force $\phi$ according to
equality ~(\ref{A.9}), we derive to the following form of the Lagrangian
(\ref{13}):
\begin{equation} {\cal L} =\left
({\dot\eta}^2/2+\bar\psi\dot\psi-\phi^2/2 \right )+\left
(-\widetilde{V}'\{\eta\}\phi+\bar\psi\widetilde{V}''\{\eta\}\psi\right )+\widetilde{V}'\{\eta\}\dot\eta. \label{16}
\end{equation} The last addend can be omitted as the total time
derivative of $\widetilde{V}\{\eta\}$ and in superfield
representation (\ref{15}) the Lagrangian (\ref{16}) assumes the
canonical form: \begin{equation} {\cal L}=\int \Lambda(\Phi){\rm
d}{\bar\chi}{\rm d}\chi,\qquad \Lambda\equiv (1/2)\Phi{\bar D}
D\Phi+\widetilde{V}(\Phi). \label{17} \end{equation} Here the kinetic
superenergy of the kernel $\Lambda$ corresponds to the first bracket
of  expression (\ref{16}) and the potential superenergy
$\widetilde{V}(\Phi)$ -- to the second one. The equalities ~(\ref{A.3}),
~(\ref{A.7}), and ~(\ref{A.8}) within the framework of the $\Phi^4$-model give
\begin{equation} \widetilde{V}={1-\sigma\over
2}\Phi^2+w{1+3\sigma\over 12}\Phi^4,\label{18} \end{equation}
where the anharmonicity parameter $w>0$ emerges because in contrast to the
order parameter $\eta$ (see (\ref{5})) the superfield
(\ref{15})
cannot be scaled by the only magnitude $\eta_m$. The supersymmetry group
generators have the form
\begin{equation} D={\partial\over
\partial\bar\chi}+\chi{\partial \over \partial t},\qquad \bar
D={\partial \over \partial\chi}+\bar\chi{\partial \over \partial t}.
\label{19}
\end{equation}
The superequations of motion ensuing from the superaction extremum
condition $S\{\Phi(z)\}{=}\int \Lambda(\Phi(z)){\rm d}z$, $z\equiv
\{{\rm{\bf r}},t,\bar\chi,\chi\}$ read:  \begin{equation} (1/2)[\bar
D, D]\Phi+\widetilde{V}'\{\Phi\}=0, \qquad (1/2)[\bar D, D]\Phi\equiv
\phi+\dot{\bar\psi}\chi+\bar\chi\dot\psi+\bar\chi\chi\ddot\eta. \label{20}
\end{equation}
Projecting Eqs.~(\ref{20}) onto basis vectors $1$, $\bar\chi$, $\chi$,
$\bar\chi\chi$ of the superspace, we arrive at equations

\def\theequation{\arabic{equation}{a}} \setcounter{equation}{20} 
\begin{equation}
\ddot\eta=-\widetilde{V}''\{\eta\}\phi+\widetilde{V}'''\{\eta\}
\bar\psi\psi, \label{21a}
\end{equation}
\def\theequation{\arabic{equation}{b}} \setcounter{equation}{20}
\begin{equation}
\phi=-\widetilde{V}'\{\eta\}, \label{21b}
\end{equation}
\def\theequation{\arabic{equation}{c}} \setcounter{equation}{20}
\begin{equation}
\dot\psi=-\widetilde{V}''\{\eta\}\psi,\label{21c}
\end{equation}
\def\theequation{\arabic{equation}{d}} \setcounter{equation}{20}
\begin{equation}
\dot{\bar\psi}=
\widetilde{V}''\{\eta\}\bar\psi. \label{21d}
\end{equation} \def\theequation{\arabic{equation}}\setcounter{equation}{21}  

\noindent
The last of this set can be
obtained from (\ref{14c}), (\ref{14d}) by replacing $V_0$ by
$\widetilde{V}$ , and the relationship (\ref{21b}) gives
the definition of the force $\phi$.  Equation (21a)
is obtained by time differentiating of equation (\ref{14a}) and
substituting the derivatives $\dot\eta$, $\dot\varphi$ from ~(\ref{A.9}),
(\ref{14b}) into resultant expression. Thus, systems (14), (21)
turn out to be equivalent with accuracy to the bare potential
$V_0(\eta)$ renormalization.  However, though the equations of the
first set are symmetrical concerning the time derivative order, in
(21) this symmetry is broken due to the transformation the
fluctuation $\varphi$ to the generalized force $\phi$.
Comparing Lagrangians (\ref{13}) and (\ref{16}) shows the fact that
the above transformation provides the standard bilinear form
of superlagrangian with respect to operators $\bar D$ and $D$. Easy
to show that, if the gauge condition $D\Phi=0$ is
satisfied the Grassmannian fields $\bar\psi$, $\psi$ are suppressed
and the single combination $\chi\bar\chi$ replaces the pair of
conjugate ones $\bar\chi$, $\chi$.  In this case the kinetic
superenergy is linear relatively to the supergroup generator,
moreover both pairs -- $\eta$, $\phi$ and $\eta$, $\varphi$ are
allowable to be used as superfield components \cite{brazh17}.
However, as is shown
in Appendix, at the description of the self-organization the
behaviour of the Grassmannian fields represents entropy and,
consequently, is essential. Therefore the gauge $D\Phi=0$ is
not fulfilled, and one should prefer making use of the generalized
force $\phi $ to the fluctuation amplitude $\varphi$.

\section{Correlation technics}\label{sec:level4}

Let us introduce the supersymmetrical correlator \begin{equation}
C(z,z')\equiv \langle\Phi(z)\Phi(z')\rangle,\qquad
z\equiv\{{\rm{\bf r}},t,\bar\chi,\chi\}. \label{22} \end{equation}
According to the motion equation (\ref{20}) its bare component
$C^{(0)}(z,z')$ meeting the potential
$\widetilde{V}^{(0)}=(1-\sigma)\Phi^2/2$ satisfies equality
\begin{equation} L_{\rm{\bf k}\omega}(\chi)C_{{\bf
k}\omega}^{(0)}(\chi,\chi ')=\delta(\chi,\chi '),\qquad L\equiv (1-
\sigma)+(1/2)[\bar D,D], \label{23} \end{equation} where the
transition to the time-spatial Fourier transforms is made and the
supersymmetrical $\delta$-function $\delta(\chi,\chi
')=-(\bar\chi-\bar\chi ')(\chi-\chi ')$ is introduced. Taking into
consideration definitions (\ref{19}) and expressions $(1/4)[\bar
D,D]^2=-\omega^2$, we obtain
\begin{equation} C^{(0)}(\chi,\chi
'){=}{1{-}(\bar\chi\chi{+}\bar\chi '\chi '){+}[(1{-}\sigma){+}{\rm
i}\omega]\bar\chi\chi '{+}[(1{-}\sigma){-}{\rm i}\omega]\bar\chi
'\chi{-}\omega^2\bar\chi\chi\bar\chi '\chi '\over
(1{-}\sigma)^2{+}\omega^2}, \label{24} \end{equation}
where index $\omega$ and spatial dispersion were omitted for
brevity. Equation~(\ref{24}) is an expansion in basis components
\begin{eqnarray}
&&{\bf B}_0=-\bar\chi\chi,\qquad {\bf B}_1=-\bar\chi '\chi
',\nonumber\\ && {\bf T}=1,\qquad {\bf T}_1=\bar\chi\chi\bar\chi
'\chi ';\label{25}\\ && {\bf F}_0=\bar\chi '\chi,\qquad {\bf
F}_1=\bar\chi\chi ' \nonumber \end{eqnarray} whose functional
product satisfies the following multiplication rules:  ${\bf
B}_0^2={\bf B}_0$, ${\bf B}_1^2={\bf B}_1$, ${\bf F}_0^2={\bf F}_0$,
${\bf F}_1^2={\bf F}_1$, ${\bf B}_0{\bf T}_1={\bf T}_1$, ${\bf
B}_1{\bf T}={\bf T}$, ${\bf T}{\bf B}_0={\bf T}$, ${\bf T}{\bf
T}_1={\bf B}_1$, ${\bf T}_1{\bf B}_1={\bf T}_1$, ${\bf T}_1{\bf
T}={\bf B}_0$, and the other multiplicands are equal to zero. Thus,
${\bf B}_{0,1}$, ${\bf T}$, ${\bf T}_1$, ${\bf F}_{0,1}$ form the
closed basis, and it is convenient to expand supercorrelator (\ref{22})
in these components: \begin{equation} {\bf C}=g_+{\bf B}_0+g_-{\bf
B}_1+S{\bf T}+s{\bf T}_1+G_+{\bf F}_0+G_-{\bf F}_1. \label{26}
\end{equation} Insertion (\ref{15}) into (\ref{22}) derives to the
coefficients
\begin{eqnarray} && g_+=-\langle \phi\eta\rangle,\qquad
g_-=-\langle \eta\phi\rangle;\nonumber\\ && S=\langle
\eta^2\rangle,\qquad s=-\langle |\phi|^2\rangle;\label{27}\\ &&
G_+=-\langle \bar\psi\psi\rangle,\qquad G_-=\langle
\psi\bar\psi\rangle,\nonumber \end{eqnarray}
where the fact is
considered that according to ~(\ref{A.9}) the field $\phi$ is exclusively
imaginary. Thus, the magnitudes $g_{\pm}$ are reduced to advanced and
retarded response functions of the order parameter $\eta$ to field
$\phi$ action; $S$ and $s$ are the autocorrelators of order parameter
$\eta$ and field $\phi$, and the functions $G_{\pm}$ determine the
correlation of Grassmannian conjugated fields $\bar\psi$, $\psi$.
The statements
\begin{eqnarray} &&
g_{\pm}^{(0)}=S^{(0)}=[(1-\sigma)^2+\omega^2]^{-1},\qquad
s^{(0)}=-\omega^2[(1-\sigma)^2+\omega^2]^{-1},\nonumber\\ &&
G_{+}^{(0)}=[(1-\sigma)+{\rm i}\omega]^{-1},\qquad
G_-^{(0)}=[(1-\sigma)-{\rm i}\omega]^{-1} \label{28} \end{eqnarray}
are valid for the bare supercorrelator (\ref{24}). Hence, according
to ~(\ref{A.9}) the relation $\langle \eta\varphi\rangle_0{=}G_-^{(0)}$ is
deduced that is a special case of the Ward identity \cite{14}. It
means that the correlator of the Grassmannian fields is reduced to
response function of the order parameter $\eta$ to the fluctuation
$\varphi$.

Expansion (\ref{26}) allows us to treat  supercorrelator (\ref{22})
as the vector of the direct product of superspaces.  Since the Fermi
components ${\bf F}_{0,1}$ do not couple with the Bose ones ${\bf
B}_{0,1}$, ${\bf T}$, ${\bf T}_1$, we may use more compact basis,
passing on from the field $\phi$ to the fluctuation $\varphi$. To
doing so, we will neglect the correlation of the self-consistent
field $\phi$ whose structure factor is $s\sim\omega^2$,
when $\omega\to 0$. Moreover, we will replace the response functions
$g_{\pm}$ to  field $\phi$ by corresponding functions $G_{\pm}$ for
the fluctuation $\varphi$.  Then the expansion (\ref{26}) takes such a
compact form \begin{equation} {\bf C}=G_{+}{\bf A}+G_-{\bf B}+S{\bf
T},\label{29} \end{equation} where the basis operators ${\bf
A}\equiv{\bf B}_0+{\bf F}_0$, ${\bf B}\equiv{\bf B}_1+{\bf F}_1$ are 
introduced that satisfy the multiplication rules:  ${\bf A}^2={\bf
A}$, ${\bf B}^2={\bf B}$, ${\bf B}{\bf T}={\bf T}$, ${\bf T}{\bf
A}={\bf T}$ (other multiplicands are equal to zero).

In the issue the behavior of self-organizing system is described
by the Lagrangian
\begin{equation}
L=(\varphi\dot\eta+\bar\psi\dot\psi-\varphi^2/2)+(\widetilde{V}'\{\eta\}
\varphi+\bar\psi\widetilde{V}''\{\eta\}\psi).\label{30}
\end{equation}
We introduce the quenched disorder
\begin{equation}
p^2={\overline{(\phi({\rm{\bf r}})-\bar\phi)^2}-(\Delta\varphi)^2\over
(\Delta\varphi)^2}. \label{31} \end{equation} Its magnitude
characterizes the field random scattering $\phi({\rm{\bf r}})$
(the dash in (\ref{31}) stands for averaging over coordinate ${\rm{\bf r}}$,
the $(\Delta\varphi)^2\equiv |\varphi_{\omega=0}|^2$ is the
square-mean fluctuation). If the quenched disorder is brought in the
action, meeting the Lagrangian (\ref{30}), component squared in
$\varphi$ fluctuation takes the form \begin{equation} -{1\over
2}\int|\varphi_{\omega}|^2{{\rm d}\omega\over 2\pi}-{p^2\over
2}\int\delta(\omega)|\varphi_{\omega}|^2{\rm d}\omega. \label{32}
\end{equation}
Here we have neglected an integration over ${\rm{\bf r}}$ and passed
to Fourier transform over frequency $\omega$.  At the equilibrium
disorder the field scatter $\phi({\rm{\bf r}})$ is reduced to the
square-mean fluctuation $(\Delta\varphi)^2$ so, that $p=0$ and
expression (\ref{32}) has the canonical form $-(1/2)\int\varphi^2{\rm
d}t$. In the case of quenching one has $p>0$ and second addend in
(\ref{32}) leads to renormalization of the bare supercorrelator (\ref{24})
whose component $S^{(0)}$ gets the
multiplier $1+2\pi p^2\delta(\omega)$ in (\ref{28}). Respectively, we
find the  operator ${\bf L}$ in the motion equation
(\ref{23}) as follows: \begin{equation} {\bf{L}}=L_{+}{\bf
A}+L_{-}{\bf B} +L{\bf T};\qquad L_{\pm}=(1-\sigma)\pm {\rm
i}\omega,\qquad L=-\left[1+2\pi p^2\delta(\omega)\right].  \label{33}
\end{equation}

To obtain equation that defines the supercorrelator (\ref{22}) one
should multiply (\ref{20}) by $\Phi(z')$ and average the result
over distribution $P\{\Phi\}$ from (\ref{12}). Thus, we get the
Dyson superequation \begin{equation} {\bf C}^{-1}={\bf
L}-{\bf\Sigma}, \label{34} \end{equation} where within the framework
of $\Phi^4$-model (\ref{18}) the self-energy superoperator
${\bf\Sigma}$ is defined by equality
\begin{equation}
\Sigma(z,z')=(2/3)w^2(1+3\sigma)^2(C(z,z'))^3, \quad z\equiv
\{{\rm{\bf r}},t,\bar\chi,\chi\}. \label{35} \end{equation} Herein
the $w>0$ is the anharmonicity parameter related to temperature $T$
and the condition $\int C(z,z) {\rm d} z=0$ is taken into account
that follows from Eqs.~(\ref{25}) and (\ref{29}).

If the anharmonicity is omitted, ${\bf\Sigma}=0$, the components
(\ref{28}), diverging at the point of transition into self-organization
state ($\sigma=1$), are obtained from Eq.~(\ref{34}). Thus, the
supersymmetrical field approach allows us to reproduce the main result
following from the Lorentz system ($6$) by means of the linear
approximation. In the general case the self-energy superfunction
should be expanded similar the supercorrelator (\ref{29}):
\begin{equation} {\bf\Sigma}=\Sigma_{+}{\bf A}+\Sigma_{-}{\bf
B}+\Sigma{\bf T}. \label{36} \end{equation} Then, according to
Eq.~(\ref{33}) the superequation (\ref{34}) is reduced to components

\def\theequation{\arabic{equation}{a}} \setcounter{equation}{36} 
\begin{equation} {G_{\pm}}^{-1}=[(1-\sigma)\pm {\rm
i}\omega]-\Sigma_{\pm}, \label{37a}
\end{equation}
\def\theequation{\arabic{equation}{b}} \setcounter{equation}{36}
\begin{equation}
S=[1+2\pi
p^2\delta(\omega)+\Sigma] G_{+}G_{-}.  \label{37b} \end{equation} 
\def\theequation{\arabic{equation}}\setcounter{equation}{37}  

\noindent
The explicit form of the expansion coefficients (\ref{36}) is given by
expression (\ref{35}). In accordance with \cite{17} the
supercorrelators multiplication should be understood in usual meaning,
not in functional one:  $(T(\chi,\chi '))^2=T(\chi,\chi ')$,
$A(\chi,\chi ')T(\chi,\chi ')=T(\chi,\chi ')A(\chi,\chi
')=A(\chi,\chi ')$, $B(\chi,\chi ')T(\chi,\chi ')=T(\chi,\chi
')B(\chi,\chi ')=B(\chi,\chi ')$, and the other multiplicands are
equal to zero.  As a result, for the spatially homogeneous case
we obtain from (\ref{35}) the following

\def\theequation{\arabic{equation}{a}} \setcounter{equation}{37} 
\begin{equation}
\Sigma_{\pm}(t)=2w^2(1+3\sigma)^2S^2(t)G_{\pm}(t),\label{38a}
\end{equation}
\def\theequation{\arabic{equation}{b}} \setcounter{equation}{37}
\begin{equation}
\Sigma(t)=(2/3)w^2(1+3\sigma)^2S^3(t).\label{38b} \end{equation} 
\def\theequation{\arabic{equation}}\setcounter{equation}{38}  

\noindent
At insertion of these expressions into the
Dyson equation (37) we will be in need of their frequency
representation that contains convolutions.  To avoid such a
difficulty let us use the fluctuation-dissipation theorem \cite{14,17}
\begin{equation} S(\omega ')=(2/\omega '){\rm Im}G_{\pm}(\omega
'),\qquad \Sigma(\omega ')=(2/\omega '){\rm Im}\Sigma_{\pm}(\omega
'),\label{39} \end{equation} where  $\omega '$ is the real
frequency. Using the spectral representation of the complex frequency
$\omega$ and integrating equalities (\ref{39}), we find
\begin{equation} S(t\to 0)=G_{\pm}(\omega\to 0),\qquad \Sigma(t\to
0)=\Sigma_{\pm}(\omega\to 0).\label{40} \end{equation} Since the
$G_{\pm}(\omega\to 0)$ gives $\chi$ in the hydrodynamical limit
$\omega\to 0$, we have  

\def\theequation{\arabic{equation}{a}} \setcounter{equation}{40} 
\begin{equation} S(t\to 0)=\chi\equiv
G_{\pm}(\omega\to 0),\label{41a}
\end{equation}
\def\theequation{\arabic{equation}{b}} \setcounter{equation}{40}
\begin{equation}
\Sigma_{\pm}(\omega\to 0)=(2/
3)w^2(1+3\sigma)^2\chi^3,\label{41b} \end{equation} 
\def\theequation{\arabic{equation}}\setcounter{equation}{41}  

\noindent
where the  expression (\ref{38b}) is used in
(\ref{41b}).  In contrast to \cite{18} here the self-energy
components $\Sigma_{\pm}$ contain only the second order of the
anharmonicity $w$.

The equations ($37$), ($38$), and  ($41$) describe the behavior of
the self-organizing system completely.  Particularly, they represent
not only the ordering phenomena but the effects of ergodicity
breaking and memory appearance as well. These effects manifest
themselves in elongation of correlators~\footnote{$^)$$\ $ Let us
point out the reverse sign of the irreversible response $\Delta$ as
compared with the definition given for thermodynamical systems, where
the ordering corresponds to low values of the noise intensity
(temperature).}$^)$ \begin{equation}
G_-(\omega)=-\Delta+G_{-0}(\omega),\qquad S(t)=q+S_0(t) \label{42}
\end{equation} at the expanse of the Edwards-Anderson memory
parameter $q=\langle\eta (\infty)\eta(0)\rangle$ and irreversible
response $\Delta=\chi-\chi_0$ that is equal to difference of
microscopic susceptibility $\chi\equiv G_{-0}(\omega=0)$ and
macroscopic magnitude $\chi_0\equiv G_-(\omega=0)$.~\footnote{$^)$$\
$ The unified function $G_-(\omega)$ can be used to define
susceptibilities $\chi_0$, $\chi$ considering that magnitudes
$\chi_0{\equiv} G_{-}(\omega{=}0)$ and $\chi\equiv G_-(\omega\to 0)$
correspond to equilibrium and out-of-equilibrium values.  In so doing
we should equip all correlators in Eqs.~(\ref{40}) and ($41$) with
index $0$ and set $\omega=0$.}$^)$

Now, let us insert the elongated correlators (\ref{42}) into expressions
($38$). Then the renormalized components of the self-energy
superfunction read  

\def\theequation{\arabic{equation}{a}} \setcounter{equation}{42} 
\begin{eqnarray} 
&&\Sigma_{\pm}(t)=2w^2(1+3\sigma)^2q^2(-\Delta+G_{\pm
0}(t))+\Sigma_{\pm 0}(t),\nonumber\\ && \Sigma_{\pm 0}(t)\equiv
2w^2(1+3\sigma)^2S_0(t)G_{\pm 0}(t)(2q+S_0(t)); \label{43a}
\end{eqnarray}
\def\theequation{\arabic{equation}{b}} \setcounter{equation}{42}
\begin{eqnarray}
&&\Sigma(t)=(2/
3)w^2(1+3\sigma)^2q^2(q+3S_0(t))+\Sigma_0(t), \nonumber\\ &&
\Sigma_0(t)\equiv(2/ 3)w^2(1+3\sigma)^2{S_0}^2(t)(3q+S_0(t)).
\label{43b} 
\end{eqnarray} 
\def\theequation{\arabic{equation}}\setcounter{equation}{43}  

\noindent
Here the nonlinear terms with respect to
the correlators $G_{\pm 0}$, $S_0$ are gathered in addends $\Sigma_{\pm 0}$,
$\Sigma_0$, in the second equality of (\ref{43a}), the relationship
(\ref{41a}) is considered, and in equations (\ref{43b})  the addends
containing $S_0\Delta\simeq 0$ are omitted. If the memory is absent
the first addends of $\Sigma_\pm(t)$, $\Sigma(t)$ vanish.  Inserting
the Fourier transforms of equations (\ref{42}), (\ref{43b}) into the
Dyson equation (\ref{37b}), within $\omega$-representation we obtain
the following relations:
\begin{equation} q_0\left
[1-(2/3)w^2(1+3\sigma^á)^2\chi_0^2q_0^2\right
]=p^2\chi_0^2,\label{44} \end{equation} \begin{equation}
S_0={(1+\Sigma_0)G_+G_-\over 1-2w^2(1+3\sigma)^2q^2G_+G_-}.\label{45}
\end{equation} The first one corresponds to the
$\delta$-like addend caused by the presence of memory, when
$\omega=0$, and the second one meets the frequencies $\omega\to 0$.
When $\omega=0$ the characteristic combination $G_+G_-$ goes to
$G_+G_-=\chi^2_0$ and the pole of the structure factor (\ref{45})
\begin{equation} 2w^2(1+3\sigma^c)^2q^2_0=\chi_0^{-2}\label{46}
\end{equation} determines the point of ergodicity breaking
$\sigma^c$, where $\chi_0=\chi$, $q_0=q$.

Now let us insert the Fourier transform of expression (\ref{43a})
into equation ($37{\rm a}$). With the help of equality (\ref{41b}) we
obtain the following equation \begin{equation}
G_-^{-1}+2w^2(1+3\sigma)^2q^2G_-=[(1-\sigma)-{\rm
i}\omega]-2w^2(1+3\sigma)^2\chi^2(q+\chi/3) \label{47}
\end{equation} for the retarded Green function in the hydrodynamical
limit $\omega\to 0$, where the relationship (\ref{41a}) is taken into
account. Hence the expression springs out for the microscopical
susceptibility $\chi{\equiv} G_-(\omega{\to} 0)$ \begin{equation}
1-(1-\sigma)\chi+(2/3)w^2(1+3\sigma)^2\chi\left [(\chi+q)^3-q^3\right
]=0.  \label{48} \end{equation}

The behavior of self-organizing system with quenched disorder is
determined completely by equations (\ref{44}), (\ref{46}), and
(\ref{48}). Herewith one should contradistinguish the macroscopical
magnitudes
$q_0$, $\chi_0$ and microscopical ones $q$, $\chi$ (the first
correspond to frequency $\omega=0$ and the second to limit
$\omega \to 0$).  The distinctive feature of this hierarchy is that
macroscopical values $q_0$, $\chi_0$ depend exclusively on quenched
disorder intensity $p$ while the  microscopical ones $q$, $\chi$ --
on thermal disorder $\sigma$.  Thereafter, to determine values $q_0$,
$\chi_0$ the magnitude $\sigma$ should be considered to be equal
to value $\sigma^c(p)$ on the line of ergodicity breaking, and the
quenched disorder intensity takes the critical value $p^c(\sigma)$
for the defining of  $q$, $\chi$. Hence, equations (\ref{44}),
(\ref{46}) determine the macroscopical values $q_0$, $\chi_0$ and
equation (\ref{48}) defines  microscopical ones $q$ and $\chi$.
Herewith the external addition $\phi_{ext}$ to the self-consistent
field $\phi$ (hereinafter $\phi_{ext}=0$ is supposed), the dispersion
$p$ of field $\phi$ fixing the intensity of the quenched disorder
(\ref{31}), and the parameter of the thermal disorder $\sigma\equiv
S^0/S_c$ act as the state parameters.

Combining the equalities (\ref{44}), (\ref{46}), we get the expression
for the macroscopical memory parameter \begin{equation}
q_0=(3/4)^{1/3}(1+3\sigma^c)^{-2/3}(p/w)^{2/3}, \label{49}
\end{equation} which increases with $p$ growth. Inserting Eq.~(\ref{49})
into Eq.~(\ref{46}) gives the macroscopical
susceptibility \begin{equation}
\chi_0=2^{1/6}3^{-1/3}w^{-1/3}(1+3\sigma^c)^{-1/3}p^{-2/3}
\label{50} \end{equation} that decreases with $p$ growth.

Fixing the memory parameter in  equation (\ref{48}) by expression
(compare it with (\ref{49}))
\begin{equation}
q=(3/4)^{1/3}(1+3\sigma)^{-2/3}(p^c/w)^{2/3}, \label{51}
\end{equation}
we find microscopical values $q$, $\chi$ dependence on
$\sigma$ (Figs.~1,~2). The fact that, when the thermal disorder is
small the functions $q(\sigma)$ and $\chi(\sigma)$ are two-valued is
distinctive for these dependencies. Since the susceptibility should
increase near the ordering point the values $q$ and $\chi$, shown in
Figs.~1,~2 by the dotted lines, are nonstable. There is an abruption
at the ordering point $\sigma_c$, where ${\rm d}\chi/{\rm d} \sigma
=\infty$. The growth of anharmonicity parameter $w$ leads to the
reduction of both susceptibilities $\chi$ and $\chi_0$.

The ergodicity-breaking point is fixed by equation
\begin{equation} 3A+(A+1)^3=1+2(1-\sigma^c)p^{-2},\qquad A\equiv
2^{5/6}3^{-2/3}(1+3\sigma^c)^{1/3}w^{1/3}p^{-4/3},\label{52}
\end{equation} obtained from condition $\chi=\chi_0$ according to
equalities (\ref{48})-(\ref{51}). As is seen from Fig.~2 the
quantity $\sigma_c$ that meets to the ordering point is defined by
the maximum value $\sigma^c$ corresponding to the breaking of
ergodicity. The dependencies $\sigma^c(p)$, $\sigma_c(p)$ which
represent the phase diagram of the self-organizing system are shown
in Fig.~3. The disordered state region corresponding to small values
of $\sigma$, and the position of it's boundary does not depend on $p$,
whereas the ergodicity region is bound by small values of
$p$.~\footnote{$^)$$\ $ Albeit we review disordered state, we can
consider that by parity of reasoning with the thermodynamical systems
the microscopical susceptibility  decreases from the
maximum value $\chi(\sigma_c)$ down to zero within the ordered region.
Thus, the
ergodicity region of ordered state is bound by the dotted line like
the one shown in Fig.~3.}$^)$ Comparing Fig.~3a with 3b shows that
the behavior of self-organizing system is rather sensitive to
the anharmonicity parameter $w$ picked.  The appropriate  dependence
$\sigma_c(w)$ for the thermal disorder parameter that corresponds to
ordering, is adduced in Fig.~4. The $\sigma_c$ takes its maximum
value $\sigma_c=1$, when $w=0$, and the growth of $w$ causes the
$\sigma_c$ monotonous decrease until $\sigma_c=0$ at
$w_c=0{.}064$. The maximum value $p_m\equiv p^c(\sigma=0)$ of
the quenched disorder at the ergodicity breaking line changes with
the value $w$ according to dependence shown in Fig.~5.  When $w$ is
small, the quantity $p_m\propto w^{-1/2}$ increases infinitely and the
ergodicity region disappears with growth of anharmonicity parameter
above critical value $w_c$.

Nonergodicity parameter is defined by solution of
Eqs.~(\ref{48})-(\ref{51}).  At constant value of quenched disorder
intensity $p$ (Fig.~6a) the three regimes are possible. At small $p$
macroscopic susceptibility $\chi_0$ exceeds microscopic quantity
$\chi(\sigma)$ for every values of $\sigma$ (see  Fig.~2), and system
is always in nonergodic state. When parameter $p$ reaches the values
which exceeds threshold $p_c$ meeting the condition
$\chi_0(p)=\chi(\sigma_c)$, the nonergodicity parameter
$\Delta=\chi-\chi_0$ takes the nonzero values only within the range
$\sigma>\sigma^c$. Microscopic susceptibility $\chi$ for every values
of $\sigma$ exceeds the macroscopic one $\chi_0$ starting from value
$p_m$ that meets the condition $\chi_0(p)=\chi(\sigma=0)$ and
system is nonergodic always. At fixed value of thermal disorder
$\sigma$ (Fig.~6b) the dependence $\Delta (p)$ is defined by infinite
increase of macroscopic susceptibility $\chi_0\sim p^{-2/3}$ in the
region of weak quenched disorder $p\to 0$. As a result, nonergodicity
parameter takes the nonzero values starting from critical value
$p^c$, and increases at further growth of quenched disorder
monotonically.

Analytic representation of $\Delta(\sigma,p)$ dependence is possible
only near the line of ergodicity breaking $\sigma^c(p)$. Setting in
Eq.~(48) $0<\sigma-\sigma^c\ll\sigma^c$, $\chi=\chi_0+\Delta$,
$\Delta\ll\chi$ in first order over small values
$(\sigma-\sigma^c)/\sigma^c$, $\Delta/\chi$ in accordance with
Eq.~(\ref{50}), we find \begin{equation}
\Delta=B(p)(\sigma-\sigma^c),\qquad B\equiv\left[1+{1\over
2}\sigma^c\chi_0-{1\over 6}\left({\chi_0}\over
{q_0}\right)^2\right]\left[(1-\sigma^c)-{2\over \chi_0}+{\chi_0\over
6 q_0^2}\right]^{-1}. \label{53} \end{equation} The coefficient $B(p)$
diverges at the point corresponding to divergence of
derivative $\partial\chi/\partial \sigma$ at $\sigma=\sigma_c$ (see
Fig.~2). At fixed value of thermal disorder it is necessary to make
the expansion in $q_0^c-q_0$ in equality (\ref{50}). Then taking
into account the dependence (\ref{49}) in linear approximation, we
obtain \begin{equation} \Delta=q_0^c\left[\left(p/p^c
\right)^{2/3}-1\right], \label{54} \end{equation} where the critical
value of memory parameter $q_0^c$ meets the ergodicity breaking point
$p^c$.

\section{Flow of sand grains motion on inclined surface}\label{sec:level5}

The viscous flow of sand grains on plain trajectory $y=y(x)$ will be
shown to represent the simplest example of self-organized criticality. The
time dependencies of horizontal $\dot{x}=\partial x/ \partial t $ and
vertical $\dot{y}=\partial y/ \partial t $ velocity components, and
slope of sand surface $y^{\prime}=\partial y/ \partial x $ fix the
behavior of a system.  In the autonomous mode they satisfy the
equations of Debye relaxation  

\def\theequation{\arabic{equation}{a}} \setcounter{equation}{54} 
\begin{equation} {{\rm d}\dot{x}\over
{\rm d}t}=- {\dot{x}\over \tau_{x}}, \label{55a}
\end{equation}
\def\theequation{\arabic{equation}{b}} \setcounter{equation}{54}
\begin{equation}
{{\rm d}\dot{y}\over {\rm d}t}=- {\dot{y}\over \tau_{y}^{(0)}},
\label{55b}
\end{equation}
\def\theequation{\arabic{equation}{c}} \setcounter{equation}{54}
\begin{equation}
{{\rm d}y^{\prime}\over {\rm d}t}=
{y^{\prime}_{0}-y^{\prime}\over \tau_{S}}, \label{55c} \end{equation} 
\def\theequation{\arabic{equation}}\setcounter{equation}{55}  

\noindent where the $\tau_{x},\,\tau_{y}^{(0)}$,
$\tau_{S}$ are corresponding relaxation times. In Eqs.~(55) is
supposed that the stationary state meets the sand grains being at
rest ($\dot{x}=\dot{y}=0$), the slope $y^{\prime}_{0}\neq 0$ plays a
role of control parameter.

Since the motion along different directions is not independent  we
ought to add to Eq.~(\ref{55a}) the transverse component $f=\dot{y}/
\gamma$ of force caused by motion along $y$ axes ($\gamma$ is the
kinetic coefficient). As a result, Eq.~(\ref{55a}) takes the form
\begin{equation} \tau_{x} \ddot{x}= -\dot{x}+a^{-1}\dot{y},
\label{56} \end{equation} where $a\equiv \gamma/ \tau_{x}$.  Note,
that in accordance with diffusion equation $\dot{y}=D
y^{\prime\prime}$, where $D$ is the diffusion coefficient, the stated
force is proportional to curvature of sand surface: \begin{equation}
f=(D/ \gamma)y^{\prime\prime}.  \label{57} \end{equation} In the
stationary state, when $\ddot{x}=0$, equation (\ref{56}) solution
yields the dependence $y=ax{+}{\rm const}$ for the tangent to sand
surface. On the other side, here the friction force (\ref{57}) becomes
proportional to longitudinal component: $f=\tau_{x}^{-1}\dot{x}$.

Taking into consideration the relationship (\ref{57}) and obvious equality
${\rm d}y^{\prime}/{\rm d}t=
\dot{y}^{\prime}+y^{\prime\prime}\dot{x}$, the expression (\ref{55c})
assumes the form \begin{equation}
\tau_{S}\dot{y}^{\prime}=(y^{\prime}_{0}-y^{\prime})-
\left(\tau_{S}/D \right)\dot{y}\dot{x}. \label{58} \end{equation}
Analogously for the vertical component we obtain \begin{equation}
\tau_{y}\ddot{y}=-\dot{y}+ {\tau_{y}\over
 \tau_{x}}y^{\prime}\dot{x},\qquad {1\over \tau_{y}}\equiv {1\over
 \tau^{(0)}_{y}} \left( 1+ {y^{\prime}_{0}\over a}
{\tau^{(0)}_{y}\over \tau_{x}} \right). \label{59} \end{equation}
Here the nonlinear addends of higher order are omitted and relaxation
time $\tau_{y}$ is introduced that depends on stationary slope
$y^{\prime}_{0}$.

Equations (\ref{56}), (\ref{58}), and (\ref{59}) coincide with the
Lorentz system (4), if  the horizontal $\dot x$ and vertical
$\dot y$ velocity components respect to the order parameter $\eta$
and conjugate field $h$, and the slope $y'$ -- to the control
parameter $S$. Then the coupling constants of the system (4)
take the form \begin{equation} A_\eta=a^{-1}, \qquad
A_h=\tau_y/\tau_x, \qquad A_S=\tau_S/D. \label{60} \end{equation}
Taking into account the relationships $A_\eta\propto v\tau_x$,
$A_h\propto v\tau_y$, and $A_S\propto v\tau_S$ obtained from the
comparison with the microscopic model (\ref{1}), it is seen that the
values of phenomenological parameters $\gamma \propto D \propto
\tau_x\propto v^{-1}$ are fixed by microscopic one $v$. In the
previous section the self-organized criticality has been shown to be
realized, if $v$ exceeds the value $v_c$. It implies that avalanche
formation on the sand surface takes place spontaneously, if shear
viscosity being proportional to the relaxation time $\tau_x$, is
bounded from above by critical value.~\footnote{$^)$$\ $ Obviously, the
relationship $\gamma \propto D$ of kinetic and diffusion coefficients
represents the Einstein expression.}$^)$ This condition is similar
to transition criteria of viscous liquid into turbulent flow mode.

\section{Conclusions}\label{sec:level6}

As the above shows at fixed values of thermal and quenched disorder
intensities $\sigma$, $p$ the behavior of self-organizing system is
represented, on the one side, by order parameter field $\eta$ and
fluctuation amplitude $\varphi$ of conjugate field, and on the other
side -- by the couple of Grassmannian conjugated fields $\psi$, $\bar\psi$.
Within the microscopic representation the Bozon and Fermion gases,
interacting between themselves by means of potential $v$ in
Hamiltonian (\ref{1}), meet them. Since with transition to
self-consistent field scheme the couple of three-tail vertexes, every
meeting the $v$, forms the four-tail one whose anharmonicity
parameter $w$ is in bare superpotential (\ref{18}), it
is possible to suppose that the relation $w=v^2$ is fulfilled. Then
the expression (\ref{A.6}) for the critical value of thermal disorder
$S_c$, in which we should take into consideration the factor $w^{-1}$
(see after (\ref{18})), takes the form \begin{equation}
S_c={\varepsilon\over 2}{\tau_S\over\tau_\eta}\left ({T\over v}\right
)^2\left ({\xi\over a}\right )^2, \label{61} \end{equation} where we
pass to dimension quantities, $T$ is the noise intensity reduced
to temperature for thermodynamic systems. The obtained $S_c(v)$
dependence has the same character as in the last formula (\ref{5}).
Identifying they, we find the expression for disagreement parameter
$\varepsilon$ that provides the time irreversibility in energy balance
equation (\ref{A.2}):  \begin{equation} \varepsilon=c\left
({\tau_0\over\tau_\eta}\right )^2;\qquad c\equiv {2\over \pi^2}\left
({a\over\xi}\right )^2{\tau_\eta^2\over\tau_h\tau_S},\qquad
\tau_0\equiv {2\pi\hbar\over T}.\label{62}  \end{equation} Here the
dimension units are used, $\hbar$ is the Planck constant. The
coefficient $c$ is defined by ratio of adiabaticity parameter
$(a/\xi)^2$, expressed in terms of spatial scales, to the
corresponding value $\tau_h\tau_S/\tau_\eta^2$ in terms
of characteristic times. It is possible to believe that
$c$ is the constant, and disagreement parameter (\ref{62}) is
determined by square of ratio of quantum fluctuation time $\tau_0$ to
macroscopic time $\tau_\eta$ of the order parameter change.
Obviously, the condition $\varepsilon\ll 1$ is satisfied always.

The peculiarity of self-consistent supersymmetic scheme presented in
sections\ \ref{sec:level2}-\ \ref{sec:level4} is that it allows us to go out of framework of the
adiabatic approximation. On conditions of its applicability, $v\ll
v_c$, the critical value of thermal disorder is defined by equality
(\ref{61}).  With growth of interaction parameter $v$ the complete
suppression of disordered state takes place, when values $v$
overcomes a critical value $v_c=0{.}252$. In this end, the
effects of mutual influence of superfield components
manifest themselves substantially, and the dependences type of
represented in Figs.~1-6 would be used.

We have shown for the simplest example of sand flow on the inclined
surface (section\ \ref{sec:level5}) that above scheme represents the avalanches
formation in self-organized criticality phenomena
\cite{2}-\cite{10}. If can suppose that the effective
potential form as usually the complicated landscape in system's
configuration space \cite{5} similar to the spin glass \cite{19}.
Therefore the theory of diffusion over nodes of hierarchical tree,
formed by set of statistical ensembles of nonergodic system, would
be used for the complete description of self-organized criticality
\cite{20,21}. The above approach pretends to describe the
conditions for single avalanche formation of lowest hierarchical
level only. As is seen from Fig.~4 one can determine the two system
types:  at supercritical values of interaction parameter $v>v_c$  the
ordering process realizes independently on external conditions; in
opposite case $v<v_c$ the system passes into self-organization mode,
if the thermal disorder intensity is above the critical value
$\sigma_c$.  Examination of the simplest sandpile model shows that on
the macroscopic level the condition of spontaneous avalanches
formation is similar to turbulence criteria.

Phase diagram of a system defines the total landscape pattern and
allows to predict the self-organized criticality character as a
function of external conditions. So, Fig. 3 shows
that system is nonergodic if the quenched disorder intensity
exceeds the maximum value $p_m$ fixed by value of $v$. According to
dependence represented in Fig.~5 at supercritical value $v>v_c$ the
self-organized criticality process does not requires the quenched
disorder at all. It is worthwhile to note that numerical
experiments Ref.~\cite{2}-\cite{6}, where
the avalanches intensities are stochastic values, meet the
large values of quenched disorder.

\section{Appendix}\label{sec:level7}
\def\theequation{{A}.\arabic{equation}} \setcounter{equation}{0}

To give  the form of synergetic system (6) for the field equations (14)
we multiply Eq.~(\ref{14c}) by $\bar\psi$ to left-hand
side, and Eq.~(\ref{14d}) by $\psi$ to right-hand side, and add the
results.  Then for quantities \begin{equation} S=\bar\psi\psi,\qquad
{\rm{\bf j}}=(\nabla\bar\psi)\psi-\bar\psi\nabla\psi \label{A.1}
\end{equation} we obtain the continuity equation
$\dot{S}+\nabla{\rm{\bf j}}=0$, that, obviously, expresses the entropy
conservation law for conserved systems. Respectively, combinations 
(\ref{A.1}) of Grassmannian fields determine the entropy $S$ and its
current $\rm{\bf j}$. It is characteristic that Eqs.~(\ref{14c}),
(\ref{14d}) for Grassmannian conjugated fields $\psi$, $\bar\psi$
differ only by sign in front of the time derivative, so that
dependencies $\psi(t)$, $\bar\psi(t)$ coincide at it inversion.
Namely this circumstance provides the entropy conservation condition
albeit for each of the fields $\psi$, $\bar\psi$ this condition is
not fulfilled: according to (\ref{14c}) in the homogeneous case the
quantity $\psi(t)$ decreases exponentially with decrement
$t^{-1}\int\limits^{t}_{0} {V_0}''(\eta(t')){\rm{d}}t'$, and the
conjugate field $\bar\psi(t)$ increases with the equal value of
increment. For the entropy $S\equiv \bar\psi\psi$ the pointed out
processes are compensated, and magnitude of $S$ is conserved. As a
result, the obtained continuity equation does not contain feedback
with order parameter $\eta$.

To switch on this feedback we ought to take into
account the disagreement of right-hand sides of Eqs.~(\ref{14c}),
(\ref{14d}) that reflects the macroscopic time irreversibility.
In this aim we introduce the coefficient $1+\varepsilon$ in
the right-hand side of Eq.~(\ref{14c}) defined by the disagreement
parameter $\varepsilon\ll 1$ (its value is determined in 
section\ \ref{sec:level6}).
Then in the right-hand side of continuity equation the term
$-\varepsilon{V_0}''(\eta)S$ appears. In addition we take into
consideration that self-organization process realizes only at
stationary current $\rm{\bf j}$ which leads to
thermostat entropy $S$ increase with constant velocity
$-\nabla\rm{\bf j}\equiv (\tau_{\eta}/\tau_S)S^0$ (at that the
entropy of self-organizing system $\Delta S\equiv S^0-S$  decreases
naturally).
As a result, entropy balance equation reads
\begin{equation} \dot{S}=(\tau_{\eta}/\tau_{S})S^0-\varepsilon
{V_0}''(\eta)S.  \label{A.2} \end{equation} From here in the
stationary regime $\dot{S}=0$, the equality $S=({\tau_{\eta}/
\varepsilon\tau_S}){S^0/{V_0}''(\eta)}$ is obtained that is reduced
to the form (\ref{7}) for the bare potential \begin{equation}
V_0=\eta^2/2+\eta^4/12.  \label{A.3} \end{equation}

Now let us examine equation (\ref{14b}) for the fluctuation field
$\varphi({\rm{\bf r}},t)$. In contrast to conjugate field $h$
the fluctuational one $\varphi$ is the nonhomogeneous even in
the stationary state. We use the approximation
$\nabla^2\varphi=(\xi/a)^2\varphi$, where $a$ is the scale
of stationary fluctuation variation, $\xi>a$ is the correlation
length to be the scale of the coordinate ${\rm{\bf r}}$.
Then in the stationary
state $\dot\varphi=0$ from Eq.~(\ref{14b}) the relationship follows
\begin{equation} \varphi={{V_0}'''(\eta)S\over
(\xi/a)^2-{V_0}''(\eta)},\label{A.4} \end{equation} showing that the
values $\varphi$, $S$ are connected at expanse of anharmonicity
${V_0}'''(\eta)\ne 0$ of bare potential only. In addition the
stability condition $\varphi>0$ requires that the scale of
nonhomogeneity does not exceed the value $({V_0}''(\eta))^{-1/2}\xi$.
At much rigorous requirement $(a/\xi)^2{V_0}''(\eta)\ll 1$
the addend $-{V_0}''(\eta)$ in the denominator may be
omitted. The pointed out inequality meets the adiabatic condition
that, however, relates not the time but spatial scales.

In terms of adiabatic approximation Eq.~(\ref{14b}) for amplitude
$\varphi (t)$ of most probable fluctuation  takes the form
\begin{equation} \dot\varphi =-(\xi/a)^2\varphi +{V_0}'''(\eta)S.
\label{A.5} \end{equation} Taking into consideration  $S(\eta)$
dependence in the stationary state, $\dot\varphi =0$, we obtain \\ 
$\varphi = (\sigma/2){V_0}'''(\eta)/ {V_0}''(\eta)$, $\sigma\equiv
{S^0/S_c}$, where the characteristic value of control parameter is
introduced \begin{equation} S_c\equiv (\varepsilon/
2)(\tau_S/\tau_{\eta})(\xi/a)^2.  \label{A.6} \end{equation} Using
approximation (\ref{A.3}) we see that found dependence
$\varphi(\eta)$ is reduced to (\ref{7}).

Lastly, let us consider equation (\ref{14a}) for the order parameter
field $\eta({\rm{\bf r}},t)$. Inserting there the
dependence $\varphi(\eta)$, we arrive at the
Ginzburg-Landau-Khalatnikov equation \begin{equation} \dot\eta
-\nabla^2\eta =-{\partial \widetilde V/\partial\eta},\qquad
\widetilde V(\eta)\equiv V_0(\eta)-(\sigma/ 2)\ln{{V_0}''(\eta)}
\label{A.7} \end{equation} that differs from (\ref{8}) by
gradient term appearance. For dependence (\ref{A.3}) the synergetic
potential assumes the form \begin{equation} \widetilde
V(\eta)=V(\eta)+\eta^4/12,\qquad V(\eta)\equiv (1/2)\left
({\eta}^2-\sigma \ln{(1+\eta^2)}\right ),\label{A.8} \end{equation}
differing from (\ref{8}) by the addend $\eta^4/12$.
This distinction is caused by the circumstance that the potential $V$
is defined at constant field $h$, whereas $\widetilde V$ -- at
fluctuation amplitude $\varphi$ fixed.  In other words, the first
potential is the  field $h$ function, whereas the second one depends
on fluctuation amplitude $\varphi$. The quantities $h$, $\varphi$
represent the couple of conjugated stationary state parameters (like
the volume and pressure in thermodynamics) and the synergetic
potentials $V(h)$, $\widetilde V(\varphi)$ are coupled by Legendre
transformation $\widetilde V=V-h\varphi$. The state equation
governing the $h(\varphi)$ dependence follows from the condition
$h=-{\partial \widetilde V/\partial\varphi}$. But it is made simpler
to introduce the field $\phi=-{V_0}'(\eta)+\nabla^2\eta{\equiv}
-\delta V_0\{\eta\}/\delta\eta$ that is reduced to the force $f$ in
Eq.~(\ref{9}) with accuracy to gradient addend. Then equation
(\ref{14a}) assumes the form \begin{equation}
\dot\eta=\phi+\varphi. \label{A.9} \end{equation} Comparing Eqs.~(\ref{A.9})
and ($6{\rm a}$), we find the important relationship \begin{equation}
h=\eta+(\phi+\varphi)\equiv (\eta-{V_0}'(\eta)+\nabla^2\eta)+\varphi.
\label{A.10} \end{equation} In the stationary state ($\dot\eta=0$)
the amplitude of the most probable fluctuation $\varphi=-\phi$
coincides with generalized force with accuracy to sign, and field
$h=\eta$ is reduced to the order parameter.  In the general case the
disagreement $\phi+\varphi\not= 0$ results in variation of order
parameter in the course of time and difference between the fields $h$,
$\varphi$ is conditioned by nonlinear component of generalized force
$\phi$.

Notice that obtained Eqs.~(\ref{A.7}), (\ref{A.5}), (\ref{A.2}) and the Lorentz
equations (6) coincide in their mathematical structure only. So, the
entropy balance equation (\ref{A.2}) contains the negative feedback similar
to equation (\ref{6c}), however it is expressed by addend
$-\varepsilon{V_0}''(\eta)S$ that is proportional to the entropy,
whereas the corresponding term $-A_S\eta h$ does not contain entropy.
The difference of addends ${V_0}'''(\eta)S$ in (\ref{A.5}) and  $A_h\eta S$
in (\ref{6b}), meeting the positive feedback, is less essential
(for bare potential (\ref{A.3}) they are equal at all). Finally the
field equation (\ref{A.7}) for the order parameter differs from
the Landau-Khalatnikov equation (\ref{8}) by account of spatial
nonhomogeneity only. As was noted, the physical reason for above
distinctions is that the Lorentz equations contains the field $h$
conjugate to order parameter, whereas the amplitude of most probable
fluctuation $\varphi$ is in the initial field equations (14).
Since the $h$ and $\varphi$ play a role of conjugated parameters of
system's stationary state, then developed field formalism and Lorentz
scheme are mutually supplementary approaches -- the first one is used
at fixed value of most probable fluctuation amplitude $\varphi$, and
the second one -- at fixed field $h$. Obviously, the second case is
realized much naturally (to pass to this we should
exploit the state equation (\ref{A.10})).

\newpage

\begin{center} CAPTIONS \\
to the paper "Field theory of avalanches formation"
by A.I. Olemskoi and A.V. Khomenko \end{center}

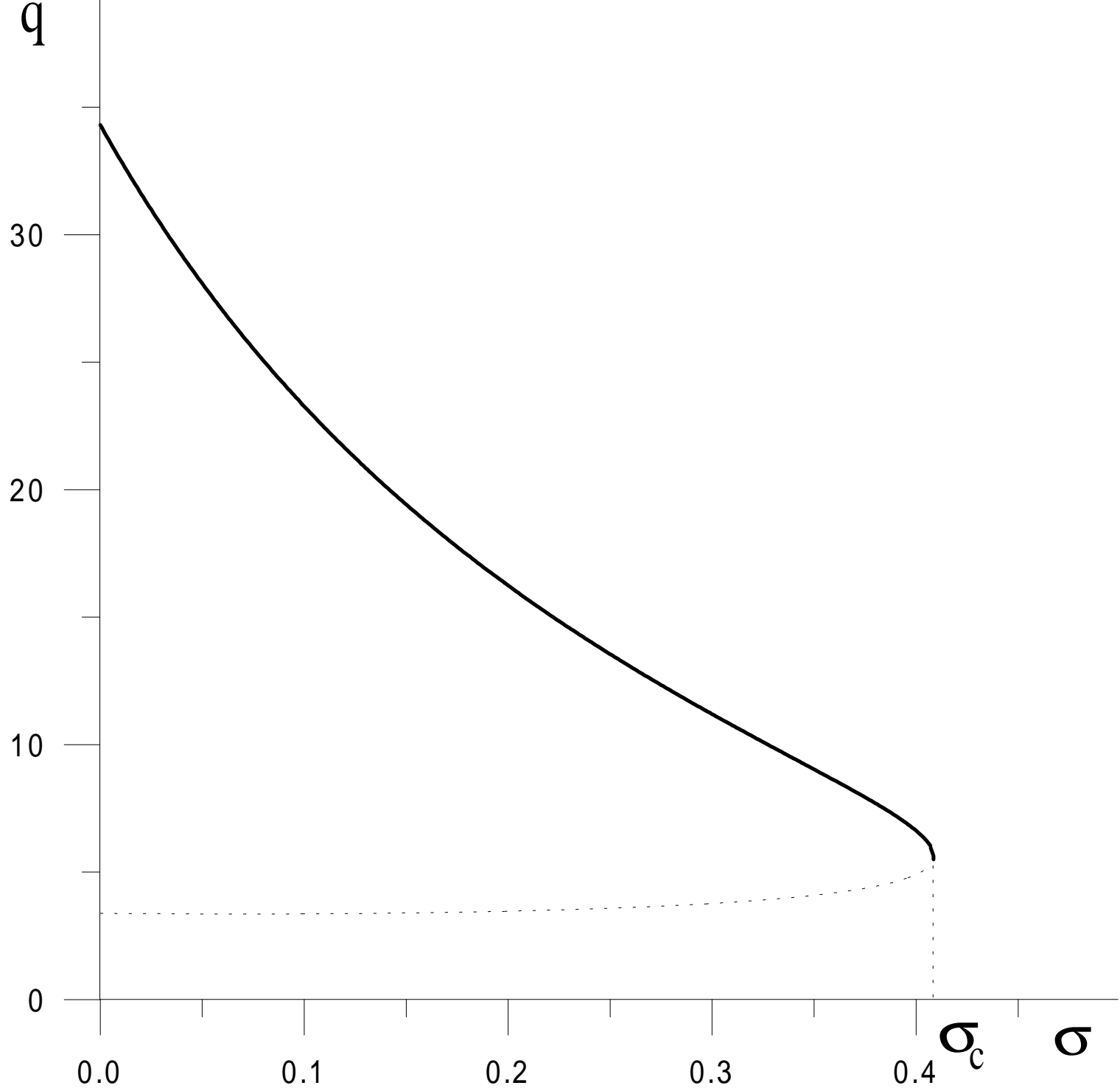
Fig. 1. Dependence of microscopic memory parameter $q$ on thermal
disorder intensity $\sigma$ at anharmonicity parameter $w=0{.}01$
(dashed line meets the unstable state, $\sigma_c$ is the ordering
point).

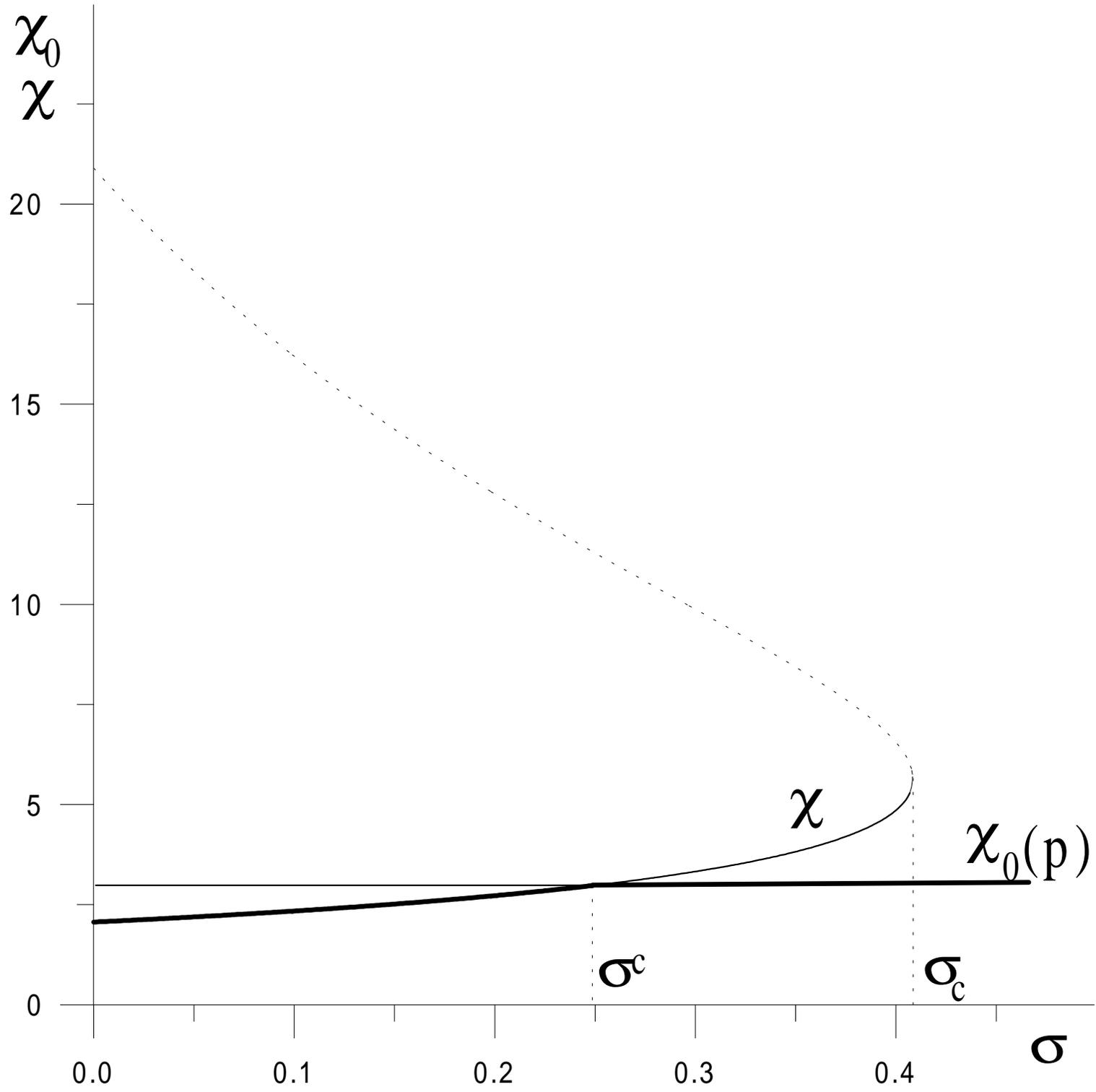
Fig. 2. Dependence of microscopic $\chi$ and macroscopic $\chi_0$
susceptibility on thermal disorder intensity $\sigma$ at
unharmonicity parameter $w=0{.}01$
(dashed and thin solid lines meet the unstable state; thick line --
stable state; $\sigma^c$ is the point of ergodicity breaking,
$\sigma_c$ is the ordering point).

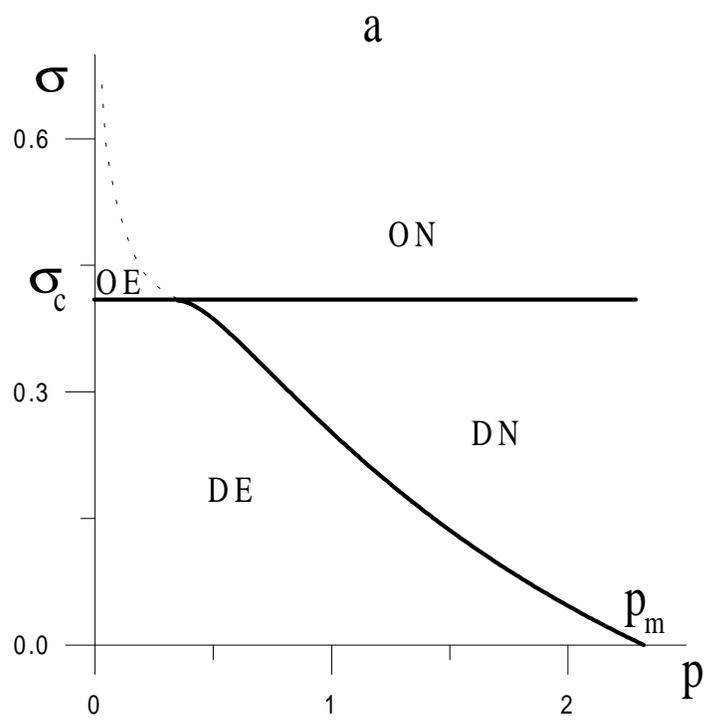
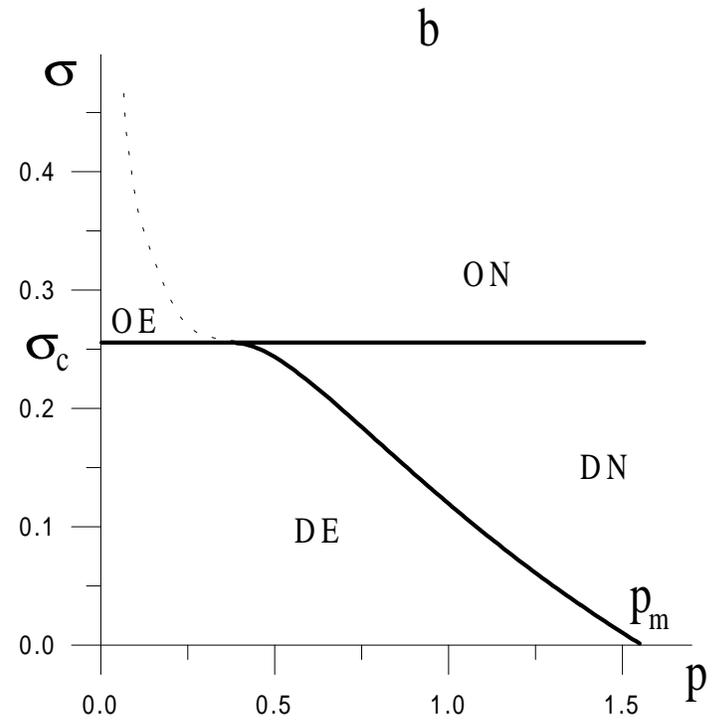
Fig. 3. Phase diagram: (a) at $w=0{.}01$; (b) at $w=0{.}02$ ($O$ is
the ordered phase, $D$ -- the disordered phase; $E$ --
ergodic, $N$ -- nonergodic; the dashed line shows the precise
boundary of ordered ergodic region).

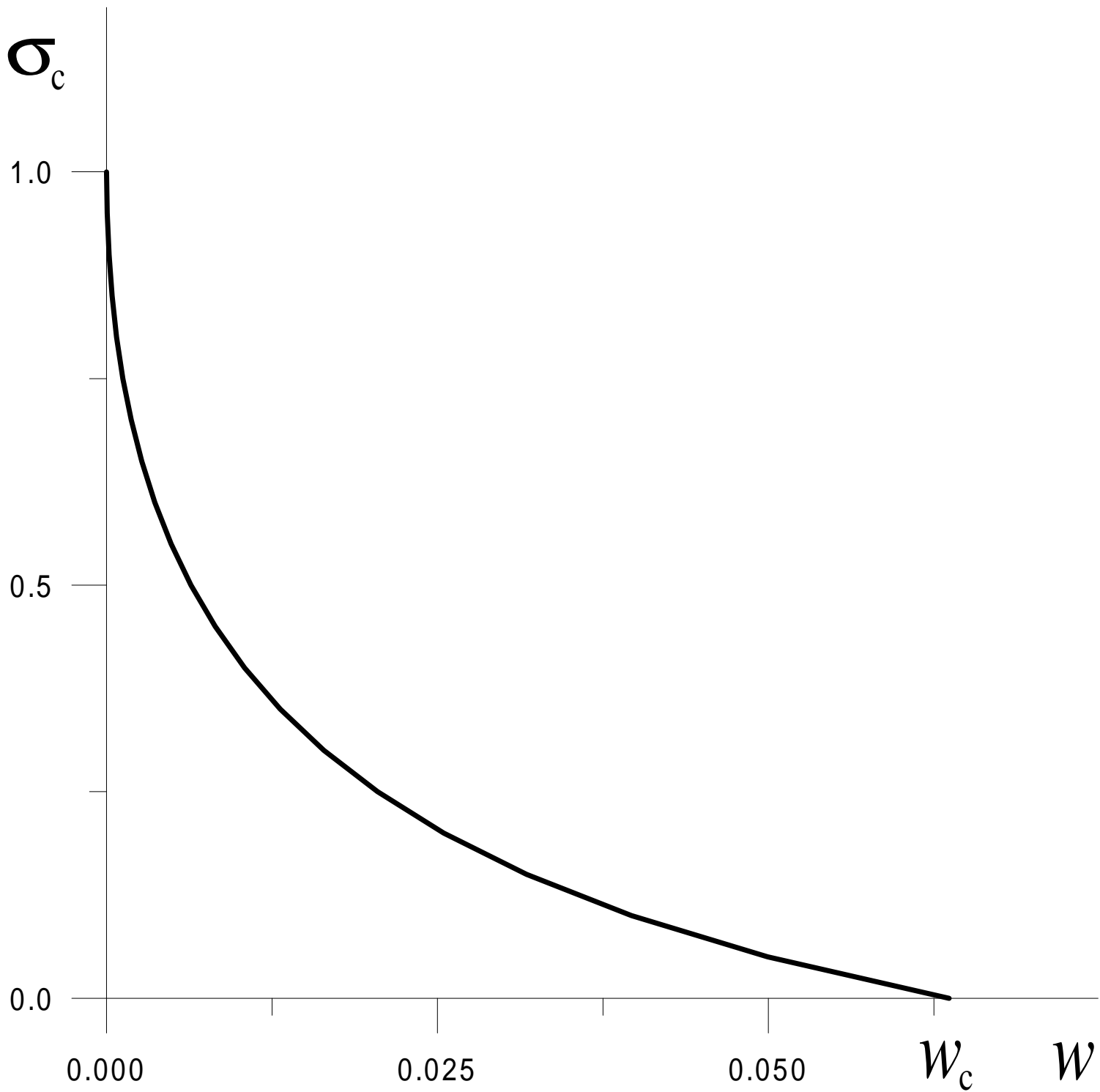
Fig. 4. Dependence of thermal disorder intensity $\sigma_c$
corresponding to the ordering point on anharmonicity parameter $w$.

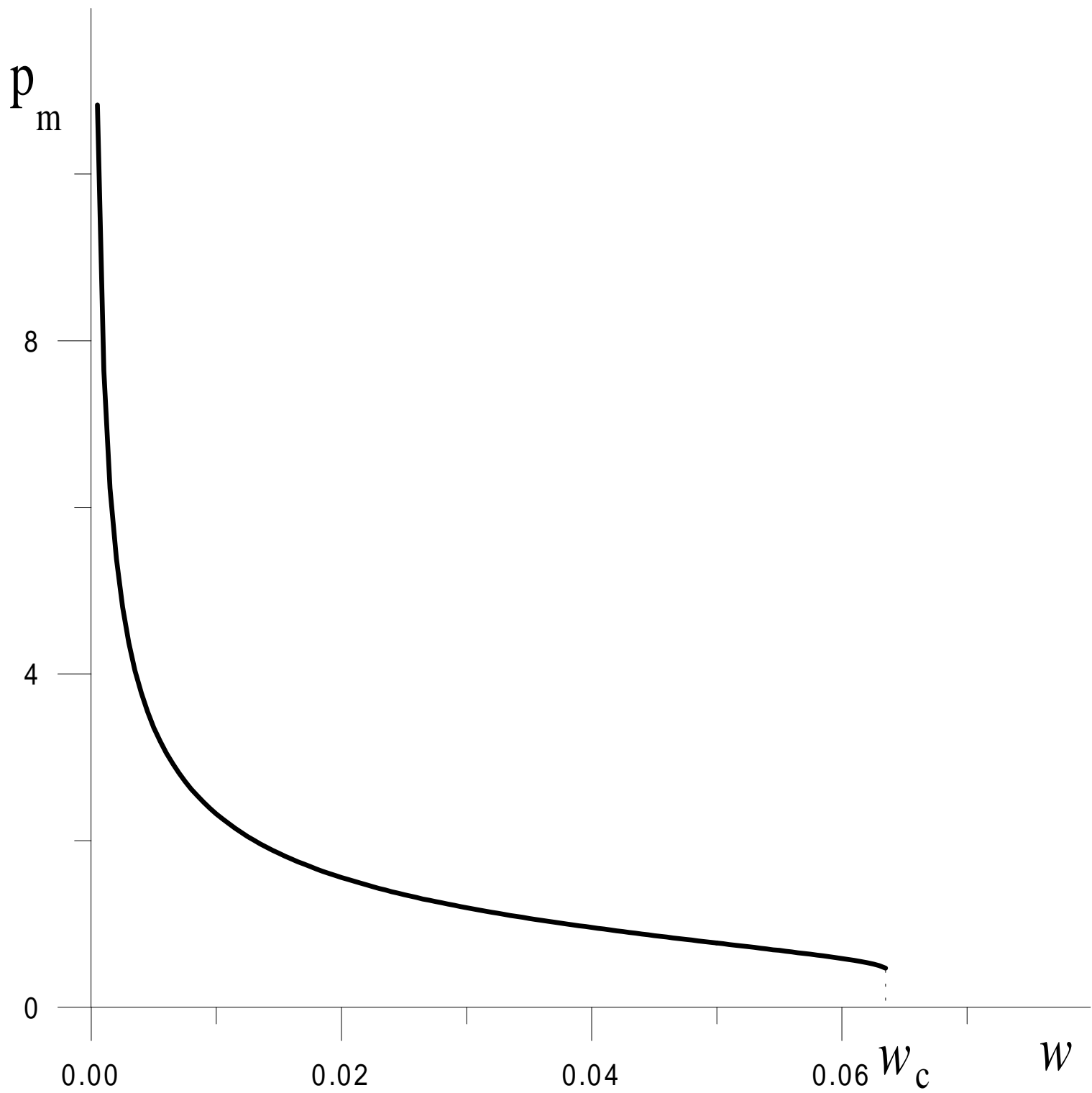
Fig. 5. Dependence of maximal intensity $p_m$ of quenched disorder at
the boundary of ergodic region on anharmonicity parameter $w$.

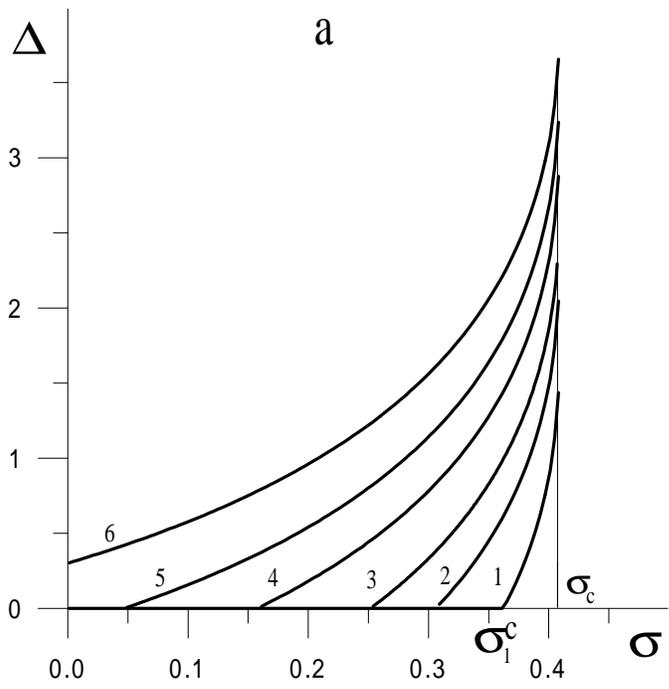
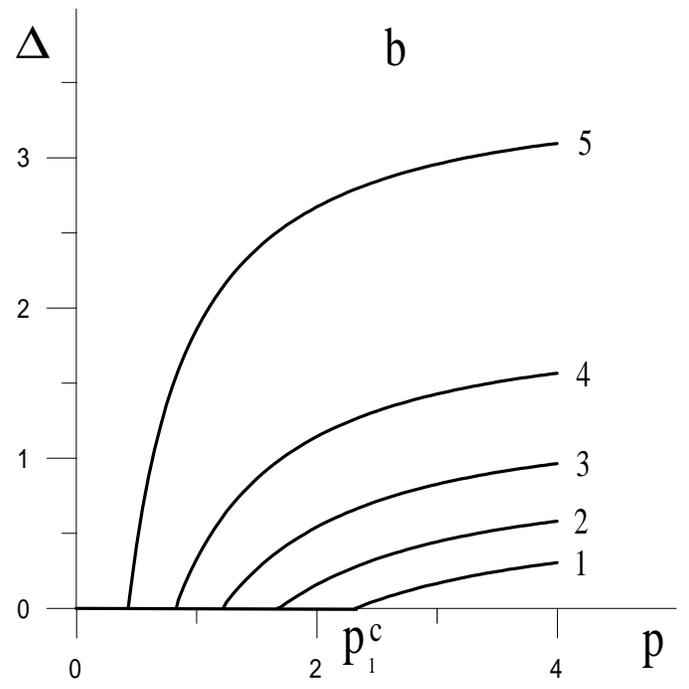
Fig. 6. Dependence of nonergodicity parameter $\Delta$:  (a) on
thermal disorder intensity $\sigma$ (curves 1-6 meet the
$p=0{.}6; 0{.}8; 1{.}0; 1{.}4; 2{.}0; 4{.}0$, respectively); (b)
on quenched disorder intensity $p$ (curves 1-5 meet
$\sigma=0{.}0; 0{.}1; 0{.}2; 0{.}3; 0{.}4$, respectively). The
anharmonicity parameter $w=0{.}01$.

\end{document}